\documentclass[12pt,a4paper]{article}
\usepackage{latexsym}

\title{\bf Muon-catalyzed fusion in "warm-fusion"}
\author{{\large K.B. Korotchenko}\\
{\small\bf Tomsk Polytechnical University, Rossia}\\
{\small\bf e-mail:  kost@phys.dfe.tpu.edu.ru}}
\date{}
\begin{document}
\maketitle

1. We would like once more to consider Ref.\cite{p1} in which
a "warm" fusion effect on $d-d $ nuclei was observed. The
experiment was carried out in Broochyven National  Laboratory
(USA). A deuterated titanium plate $TiD $ highly saturated
with deuterium was bombarded by heavy water clusters
$(D_2O)_N D^+ $ (further we shall denote them as $(D_2 O)_N )
$ with energy 300 keV.

As at is seen from Ref.\cite{p2} the results  obtained  are
in a sharp contradiction to the standard concept of these
processes. Attempts, made to explain the results of
Refs.\cite{p2}, \cite{p3} are unconvincing because they are
aimed to reduce the experimental  results to an old idea of
fusion by macroparticles collision \cite{p4}.

In this paper we represent a fusion model operating within  a
system  $(D_2 O)_N  - TiD $ which enables one to explain the
main peculiarities of the experiment \cite{p1} in detail not
using the idea  mentioned above \cite{p4}.

2. Let us formulate the initial hypothesis  which, from our
point of view, is necessary to explain the experiment
discussed :
\begin{itemize}
\item[\ ] {\bf g1.}
{\it The output $N_{off} $ of a fusion reaction (counting
rate) is proportional to the deuterium concentration $C_D $
in the cluster $(D_2 O)_N $ (the $D $ atoms concentration in
the target is constant).}
\end{itemize}

Hence, $N_{off} \sim C_D \sim N/V \sim N/R^3 $, where $N $
is the number of $D $ atoms in the cluster $(D_2 O)_N $;
$R $ is the average radius of the cluster. Thus
\begin{equation}
\label{e01}
 R \sim (N/N_{off} )^{1/3} .
\end{equation}

3. Hypothesis given above allows one to reproduce  many
peculiarities of the experimental data \cite{p1} but two
significant problems are still not clear:
\begin{itemize}
\item[\ - ]
why does a sharp change of fusion  reaction  behavior  take
place for $N \approx 110 $?
\item[\ - ]
why does the number of deuterium "supplied" by clusters for
the fusion reaction greatly decrease for $N > 500 $
(vanishing practically) ?
\end{itemize}

Moreover, the most important problem - why  does  the  fusion
reaction occur at all when the deuterium  atoms  possess such
a relatively low energy $(0.5...12 keV) $? - remains
unsolved.

We think that to solve these problems we need the following
postulate:
\begin{itemize}
\item[\ ]
{\it in a molecule $D_2 O $ (in a cluster of heavy water
$(D_2 O)_N ) $ when its specific energy is of the order of
$150 eV $ one of the molecular bonds $O - D $ is changed into
$O - D_{\mu} $ i.e. a molecule $D_2 O $ transforms into a
muonic molecule.}
\end{itemize}

4. Further considerations are sufficiently clear. First of
all, we think that in the experiment \cite{p1}, the process
of $\mu$-catalyzed nuclear fusion reaction was observed. Let
us take into consideration that the initial hypothesis {\bf
g1} merely becomes one of the characteristics of the
$\mu$-catalyzed fusion (see for example Ref.\cite{p5} ). In
this case however the formula (\ref{e01}) is to be
represented in the form
\begin{equation}
\label{e02}
R \sim \big(N (1 - N_{off} B )/( kN_{off} )\big)^{1/3} ,
\end{equation}
where $k $ and $B $ are constants. When the value of the
output $N_{off} $ of the fusion reaction is not so great,
$N_{off} B \ll 1 $. Ionization energy of a muonic atom
$D_{\mu} $ is to be equal to $T_{\mu} = 206\: T_o $, where
$T_o $ is the ionization energy of a hydrogen atom. Hence,
$T_{\mu} = 2.8 keV $. This is exactly the same energy that
muonic atom $D_{\mu} $ has in a cluster at $N = 107 $ (the
cluster energy being equal to $300 keV $). Thus the
processes, which occur in the experiment \cite{p1}, can be
described in the following way:
\begin{itemize}
\item[\ - ]
for a small cluster size $(N = 25...110) $ the specific
energy of cluster muonic atoms $D_{\mu} $ is greater than
their ionization energy; there appears a sufficiently high
local concentration of  nuclei $d $ and free $\mu$-mesons in
the target after its collision with the cluster and
"destroying"  the  latter; afterwards the $\mu$-catalyzed
fusion reaction develops according to the standard scheme;
\item[\ - ]
in large cluster $(N = 120...150) $ the specific energy of a
muonic atom $D_{\mu} $ is not sufficient to ionize it; as a
result the cluster "supplies" in the target not charged
particles ($d $ and $\mu $) but neutral muonic atoms $D_{\mu}
$; probability  of collisions  between them is very small;
hence, in this case, the fusion takes place mainly owing to
collisions of muonic atoms $D_{\mu} $  with target nuclei $d
$; consequently the relative concentration $C_D $ in
comparison with light cluster sharply decreases.
\end{itemize}

Thus, we answered the first question of part 3 about causes
of a threshold character of the fusion reaction at $N \approx
110 $. To answer the second question we note that maximal
cluster size $N $ at which the cluster still has energy for
dissociation of a muonic molecular axis $D_2 O $ (which moves
so that its molecular axis $O - D $ and the movement
direction make angle $\vartheta $ ), can be calculated by the
formula
\begin{equation}
\label{e03}
 N = T_N \cos^2 \vartheta / T_{dis} ,
\end{equation}
where $T_N $ is the cluster energy and $T_{dis} $ is the
dissociation energy of a muonic molecule $D_2 O $.

Unfortunately, the  author  doesn't  know experimental value
of the dissociation energy for a muonic molecule $D_2 O $,
but one can evaluate this energy judging from the following
considerations. For the molecule $H_2 O $ the electron
affinity is $0.9 eV $.  Hence, for the muon the electron
affinity is to be equal to 185 eV. Since, for dissociation of
a muonic molecule $D_2 O $ through the canal $D_2 O
\longrightarrow D_{\mu} + DO $ the molecular bounding~
maintained by a muon is to be broken. Because of it the
dissociation energy of a muonic molecule $D_2 O $ does differ
from that of an ordinary molecule $(i.e. 5 eV ) $ by the $185
eV $ value (the electron affinity for muon). Thus, $T_{dis}
\approx 190 eV $.

Then from (\ref{e03}) for $T = 300 keV $ and $\vartheta = 0 $
we obtain that $N = 1580 $. If we take into account the fact
that according to the classical theory of molecular bounds
the angle between molecular  axis $O - H $ in a water
molecule is to be equal to $104^\circ $ then for $T_o = 300
keV $ and $\vartheta = 52^\circ $ we obtain the $N \approx
550 $. However orientation of different molecules $D_2 O $ in
a cluster toward the movement direction is likely to be
rather random. Therefore in the range of $N $ from $550 $ to
$1580 $ the number of muonic molecules $D_2 O $ of a cluster
which energy is sufficient for dissociation is smoothly
vanishing.

In conclusion we note that the statement expressed as a
postulate in the part 3 is a quality result of the quantum
theory interpretation, which is developed by the author in
his previous paper (Refs.\cite{p6}).

\end{document}